\documentclass[final,5p,times,twocolumn]{elsarticle}

\usepackage{graphicx}

\usepackage{amssymb}

\begin{document}

\begin{frontmatter}

\title{A Site Evaluation Campaign for a Ground Based Atmospheric Cherenkov Telescope in Romania}

\author[inst1]{A.A.~Radu\corref{cor1}}
\ead{aurelian.radu@spacescience.ro}

\author[inst2]{T. Angelescu}
\author[inst3]{V. Curtef}
\author[inst4]{F. Delia}
\author[inst1]{D. Felea}
\author[inst5]{I. Goia}
\author[inst1]{D. Ha\c{s}egan}
\author[inst5]{B. Lucaschi}
\author[inst5]{A. Manea}
\author[inst1]{V. Popa}
\author[inst5]{I. Rali\c{t}\u{a}}
\author[inst4]{R. V\u{a}c\u{a}reanu}

\cortext[cor1]{Corresponding author at: Institute for Space Sciences (ISS), 409 Atomistilor St., Bucharest-Magurele, Romania. Tel.: +40 21 4574471; fax: +40 21 4575840}

\address[inst1]{Institute for Space Sciences (ISS), 409 Atomistilor St., Bucharest-Magurele, Romania}
\address[inst2]{Bucharest University, Faculty of Physics, P.O. BOX MG-012, Bucharest-Magurele, Romania}
\address[inst3]{Universit\"{a}t W\"{u}rzburg, D-97074 W\"{u}rzburg, Germany}
\address[inst4]{Technical University of Civil Engineering Bucharest, 124 Lacul Tei Boulevard, Bucharest, Romania}
\address[inst5]{National Meteorological Administration, 97 \c{S}oseaua Bucure\c{s}ti-Ploie\c{s}ti, Bucharest, Romania}

\begin{abstract}
Around the world, several scientific projects share the interest of a global network of small Cherenkov telescopes 
for monitoring observations of the brightest blazars - the DWARF network. A small, ground based, imaging atmospheric 
Cherenkov telescope of last generation is intended to be installed and operated in Romania as a component of the DWARF 
network. To prepare the construction of the observatory, two support projects have been initiated. Within the framework 
of these projects, we have assessed a number of possible sites where to settle the observatory. In this paper we submit 
a brief report on the general characteristics of the best four sites selected after the local infrastructure, the 
nearby facilities and the social impact criteria have been applied.
\end{abstract}


\end{frontmatter}

\section{Introduction}
\label{sec1}

The best way to detect very high-energy (VHE) $\gamma$-rays (100 GeV $<$ E $<$ 100 TeV), of cosmic origin, from 
the ground is by imaging the Cherenkov light produced by the secondary particles once the $\gamma$-rays interact within 
the atmosphere. This method employs elaborate instruments, as the ground based Imaging Atmospheric Cherenkov Telescopes 
(IACTs) \cite{buckley-08}. The major IACT experiments currently in operation, MAGIC \cite{MAGIC}, VERITAS \cite{VERITAS}, 
H.E.S.S. \cite{HESS}, and CANGAROO-III \cite{CANGAROO} have shown us an immense potential for scientific discoveries with 
significant consequences for astrophysics, cosmology and particle physics. A new era of outstanding precision will begin 
with the future Cherenkov Telescope Array (CTA) - the next generation of highly automated telescopes for gamma-ray 
astrophysics \cite{doro-09}.

The number of known VHE $\gamma$-ray sources has exponentially grown and is presently more than 100 \cite{naurois-09}. 
Some of them belong to the category of Active Galactic Nuclei (AGNs). It is inferred that AGNs have large luminosities, 
beamed emission and their energy source is the release of gravitational energy from an accretion disk surrounding a 
super-massive black hole \cite{albert-07}. Particularly interesting for the VHE $\gamma$-ray community are the blazars, 
a sub-class of AGNs whose relativistic plasma outflows (jets) point towards the observer. So far, more than 10 AGNs 
have been detected in the VHE $\gamma$-ray window and most of them are blazars. One distinctive feature of blazars is 
the strong variability of their VHE emission. It ranges from years down to minutes scales \cite{bretz-09}. The study of 
this variability may help us better understand the central engine of AGNs, the particle acceleration within their plasma 
jets, and the propagation of these jets. Long-term monitoring observations of the blazars and maybe some other sources 
will provide the VHE $\gamma$-ray astronomy of the future years with data of paramount importance for solving these problems.

The major IACT experiments are presently performing observations at their sensitivity limit or in multi-wavelength campaigns 
for most of their operation time. If monitoring observations take place (see MAGIC \cite{hsu-09}), the amount of time 
assigned is, by far, not sufficient.

Under these circumstances and due to the importance of the observational data, a global network of several small Cherenkov 
telescopes was proposed to be operated in a coordinated way for long-term monitoring observations of the brightest blazars - 
the DWARF (Dedicated Worldwide AGN Research Facility) Network \cite{bretz-08}. This network will have to be distributed 
around the globe for 24/7 monitoring, preferably with temporal overlap and redundancy to account for weather and duty cycle 
constraints, as well as for muon background reduction. The prototype telescope of this network is the former HEGRA CT3 
telescope, located at the Roque de los Muchachos on the Canary Island of La Palma, completely refurbished with an enlarged 
mirror area and a robotic design \cite{backes-09}.

The DWARF type telescopes have smaller reflectors (total surface $\approx$ 10 $m^{2}$) than the largest ones (MAGIC, HESS 
$>$ 100 $m^{2}$) and this limits the observations to the brightest VHE gamma-ray sources at a few hundred GeV energy 
threshold, still acceptable for monitoring purposes.

So far, around the world, several scientific projects share the interest of the DWARF network in monitoring observations 
of the brightest blazars. Since 2005, the Whipple telescope located on Mt. Hopkins in Arizona, USA has been used for nightly 
monitoring observations and since 2007 it was decided that Whipple observations will dovetail with those of the DWARF 
telescope \cite{backes-09}. The TACTIC telescope situated on Mt. Abu, India is also dedicated to long-term monitoring 
observations and it can be operated within the DWARF network \cite{koul-07}. Two of the former HEGRA air Cherenkov telescopes 
will be refurbished, installed and operated as the OMEGA project on the Volcano Sierra Negra in the state of Puebla, Mexico 
\cite{alfaro-09}. The main scientific goal of OMEGA is testing at very high altitude ($\approx$ 4100 m a.s.l.) its ability 
to detect the Cherenkov light from the air showers generated by VHE $\gamma$-rays. However, OMEGA will be also used for 
monitoring observations of the brightest blazars and to follow up candidate HAWC \cite{HAWC} sources. The Star Base Utah 
will be a stereoscopic system of two telescopes located close to Salt Lake City, USA \cite{finnegan-08}. After the Cherenkov 
cameras will be completed, the system will be used for monitoring observations, as well. CROATEA is proposed to be a small 
IACT system based on two ex-HEGRA telescopes, located near the Adriatic coast in Croatia \cite{hrupec-08}. CROATEA will serve 
as a test telescope for new photodetectors and as an observatory devoted to known AGNs.

To join the international efforts on understanding the physics of VHE $\gamma$-rays, a small, ground based, imaging atmospheric 
Cherenkov telescope of last generation is intended to be installed and operated in Romania as a component of the DWARF network.

\section{The Romanian Cherenkov telescope}
\label{sec2}

On the initiative of the Institute of Space Sciences at Bucharest-Magurele, a consortium has been established in order to 
prepare the construction, in Romania, of an observatory for VHE $\gamma$-ray studies. The observatory will employ, at the 
beginning, a small, ground-based, last generation, Cherenkov telescope. Upgrades and developments are possible in the future.

The major scientific goals to be accomplished are: to work in synergy with gamma-ray telescopes around the world for long term 
monitoring observations of the brightest blazars and eventually other VHE $\gamma$-ray sources, to perform broadband multi-wavelength 
observations in collaboration with other observatories, to participate in multi-messenger observations of $\gamma$-rays and 
neutrinos with neutrino observatories and last but not the least, to exploit its great educational potential for the students 
interested in the field.

To prepare the construction of the observatory, two support projects have been initiated. The first one is focused on building 
a dedicated instrument to measure the level of the background light of night sky (LONS) \cite{mirzoyan-94}. Within the framework 
of the second project, a site testing campaign has started at the end of 2007.

During this campaign, we have assessed all the places where middle altitude meteorological stations from the Romanian national 
network have been in operation over the last ten years. A time range of ten years is usually considered appropriate when general, 
long-term tendencies are looked for in meteorological data sets. Subsequently, taking into consideration technical, economical and 
social selection criteria (section III.b and III.c), we generated a short list of four locations (Table 1 and Figure 1). However, 
this is not the list of final choice. An upgraded version will be produced when the future in-situ light pollution measurements 
will be completed.

A statistical data analysis of the standard meteorological parameters (air temperature, humidity, air pressure, wind speed) was 
performed on the ten years data from the national meteorological database for the sites on the short list. The results will be 
published in a future paper.

In this paper, we submit a brief report on the general characteristics of the sites included on the present short list, from the 
point of view of the local infrastructure, the nearby facilities and the social impact.

\begin{table}
\caption{\label{tab1}The latitude, the longitude and the altitude of the sites on the present short list, for the settlement of a Cherenkov telescope in Romania.}
\begin{tabular}{cccc}
\hline\noalign{\smallskip}
Site & Latitude & Longitude & Altitude [m] \\
\noalign{\smallskip}\hline\noalign{\smallskip}
Baisoara & $46^{o}$ $32^{'}$ $08^{''}$ N & $23^{o}$ $18^{'}$ $37^{''}$ E & $1357$ \\
Rosia Montana & $46^{o}$ $19^{'}$ $03^{''}$ N & $23^{o}$ $08^{'}$ $21^{''}$ E & $1198$ \\
Semenic & $45^{o}$ $10^{'}$ $53^{''}$ N & $22^{o}$ $03^{'}$ $21^{''}$ E & $1432$ \\
Ceahlau & $46^{o}$ $58^{'}$ $39^{''}$ N & $25^{o}$ $57^{'}$ $00^{''}$ E & $1897$ \\
\noalign{\smallskip}\hline
\end{tabular}
\end{table}

\begin{figure}
\resizebox{1.\hsize}{!}{\includegraphics{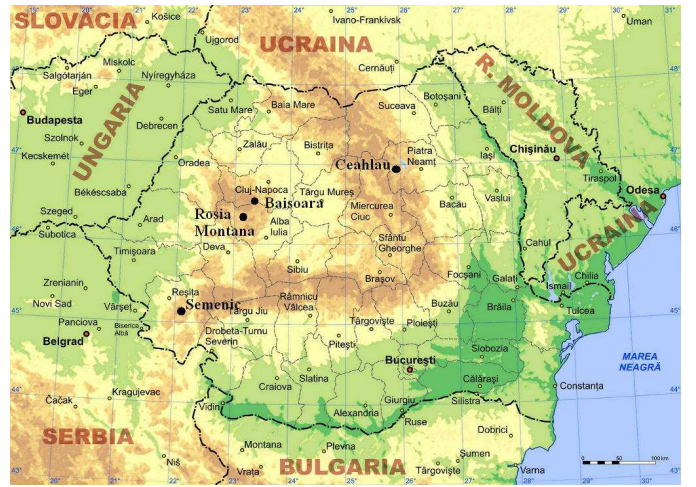} }
\caption{\label{fig1}The four candidate sites on the present short list, for the settlement of a Cherenkov telescope in Romania.}
\end{figure}

\section{General criteria for site assessment}
\label{sec3}

Systematic site testing is generally justified by the need to find the best new sites for the settlement of an astronomical 
observatory and sometimes by the need to make a proper selection among sites already in use and considered as good ones.

All the astronomical telescopes are designed and operated in close relation with the local conditions. The site has a significant 
influence on the good quality of the future astronomical data as well as on the scientific productivity of the observatory, over 
its lifetime. The characteristics of the site affect the cost and ease of construction and operation. There is also an impact on the 
activities of management, technical support, and personnel recruiting. Therefore, the selection of a site is always a critical issue.

No requirements are created for a site in the form of limits for certain parameters, as there are generally no hard cut-offs beyond 
which a site becomes unsuitable. Instead, the scientific teams measure the technical properties of the sites with the highest 
accuracy and longest temporal baseline possible. These parameters are subsequently balanced against each other using a methodology 
developed during the course of the site testing process \cite{schoeck-09}.

IACTs are mainly used as astronomical instruments, observing in the near UV bands (UVB and UVA) and the visible part of the 
electromagnetic spectrum (the wavelength range of Cherenkov light photons goes roughly from 300 to 600 nm \cite{konopelko-04}). Most 
of the criteria that apply to the selection of the sites hosting astronomical telescopes operating in this wavelength range, also 
apply to the sites hosting IACTs. A review of these criteria follows below.

\subsection{Physical criteria}
\label{sec3.1}

\begin{enumerate}[a)]
\item the site should be located at moderate to high altitudes (1000 - 2500 m), if possible, above the inversion layer \cite{merck-04}; 
it is under consideration if exceeding the altitudes of $\approx$ 3000 m can be beneficial or not to an IACT \cite{alfaro-09},
\item the site with the lowest possible value of geomagnetic field (GF) should be selected in order to minimize the influence of this 
parameter on the observatory performance \cite{commichau-08},
\item the site should be characterized by good atmospheric transparency (minimization of Rayleigh and Mie scattering). This happens 
when there is a low cloud coverage, a low amount of dust in the air and a low level of chemical pollution,
\item the site should be characterized by a low level of humidity. High humidity can produce severe damages to electronics and it 
can increase the absorption of the Cherenkov light in the blue,
\item low speed winds are required for the site as moderate speed winds can put the electronic camera in oscillation and high speed 
winds can damage the large mirror of the telescope,
\item the temperatures at the site should be moderate and preferably free from ice (protection of mirrors) and snow (reduced level 
of albedo radiation),
\item the site should be characterized by a low level of natural (e.g. aurora borealis) or man-made light pollution,
\item the atmospheric turbulence above the site should be low (stable atmosphere) in order to enable a maximum response for highly 
inclined showers \cite{barrio-98}. However, IACTs are less influenced than the optical ground based telescopes by the turbulences in 
the Earth's atmosphere, because the objects under study are VHE $\gamma$-ray induced air showers and the fluctuations in the shower 
development are considerably higher than the effects of wavefront distortions \cite{garczarczyk-06}.
\end{enumerate}

\subsection{Technical criteria}
\label{sec3.2}

\begin{enumerate}[a)]
\item	the site should be one of good geological and geotechnical conditions (seismic activity, mechanical properties of the soil, 
vibration transmission properties),
\item	the site should have good access roads for transportation purposes,
\item the site should offer a large enough area for the installation and the safe operation of the telescope and the associated 
equipment and buildings,
\item	electricity and water supplies should be available on site.
\end{enumerate}

\subsection{Economical and social criteria}
\label{sec3.3}

\begin{enumerate}[a)]
\item	the construction and operation costs associated to the site should be under the available budget,
\item	the site should allow for land ownership or eventually the rent paid should be as low as possible,
\item	the travel costs to the site should be low,
\item	the labor force availability and the economic impact of building the telescope (new jobs on the market) should be considered,
\item	the cultural, the archeological, the environmental and the land use potential restrictions have to be considered,
\item	the reasonable proximity of the site to academic centers is desirable.
\end{enumerate}

Based on the above considerations, when site selection programs are underway, systematic site-to-site comparisons are carried out 
and the site that best fits the above criteria is selected.

\section{The selected sites}
\label{sec4}

\subsection{Baisoara}
\label{sec4.1}

The site is located in the region of Transilvania, the Cluj County. There is a modernized, paved road, 18 km long, connecting the site 
to the local county road. The closest major cities are Cluj-Napoca ($\approx$ 60 km, $\approx$ 318,000 inhabitants) and Turda ($\approx$ 
55 km, $\approx$ 60,000 inhabitants). The closest bus station is 18 km away, in the village of Baisoara and this is also the place where 
basic supplies can be purchased. The closest hospital is located in Turda. Cluj-Napoca harbors the closest railway station and the closest 
airport (domestic and international connections). Rental car services are available in Cluj-Napoca. In Figure 2 we show the road map of 
the area.

\begin{figure}
\resizebox{1.\hsize}{!}{\includegraphics{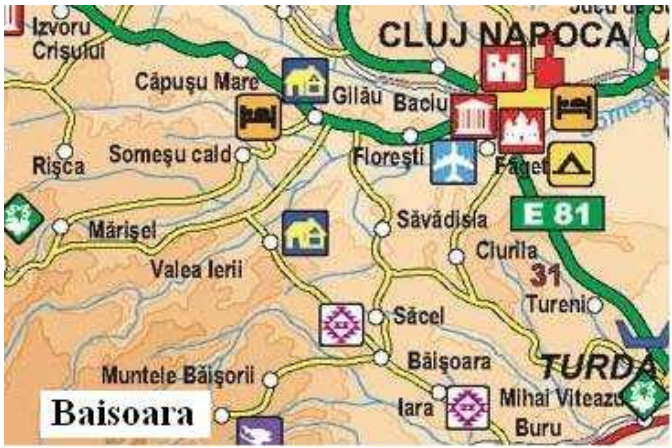} }
\caption{\label{fig2}The local road map for the site of Baisoara.}
\end{figure}

Electricity and water supply are available on the premises. The communications services are provided by a national operator (cellular 
phone and internet). Major universities with Physics and/or Mathematics departments in the region are located in the following cities: 
Cluj-Napoca, Targu Mures ($\approx$ 160 km), Baia Mare ($\approx$ 200 km) Oradea ($\approx$ 200 km), Alba Iulia ($\approx$ 220 km) and 
Sibiu ($\approx$ 220 km). Offices and laboratories associated to the observatory may be operated in Cluj-Napoca, the major academic center 
of the region. The site is not placed inside a national park (reservation area) and no archeological, cultural or social restrictions may 
impede on the settlement of the observatory. No chemical hazards can be identified in the area. The meteorological station whose database 
was used for analysis is located next to a skiing resort, this contributing some light pollution. However, the telescope is going to be 
built outside the location of the meteorological station in a more suitable, darker place in the region of Baisoara. The future in-situ 
light pollution measurements will help us to spot out the best place for the settlement of the telescope.

\subsection{Rosia Montana}
\label{sec4.2}

The site is located in the region of Transilvania, the Alba County. There is a 1.5 km paved road that connects the site to the national 
road from the city of Abrud ($\approx$ 14 km). The closest major city is Alba-Iulia ($\approx$ 80 km, $\approx$ 67,000 inhabitants). 
The closest bus station and the closest hospital are located in Abrud. The closest railway station is located in the city of Zlatna 
($\approx$ 50 km). The city of Cluj-Napoca ($\approx$ 133 km) harbors the closest airport (same connections as for Baisoara). Rental 
car services are available in Cluj-Napoca. In Figure 3 we show the road map of the area.

\begin{figure}
\resizebox{1.\hsize}{!}{\includegraphics{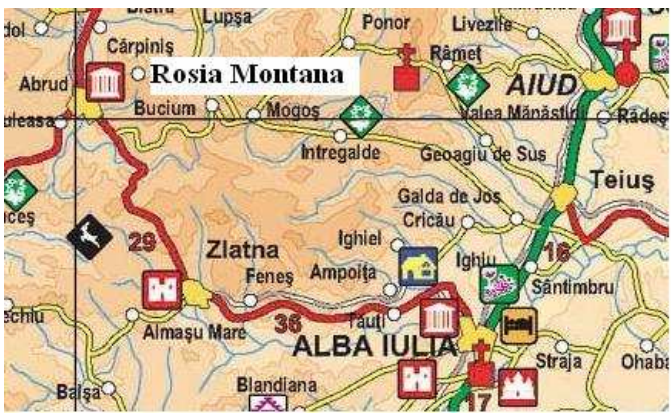} }
\caption{\label{fig3}The local road map for the site of Rosia Montana.}
\end{figure}

Electricity, water supply and sewage are available on site. The communications services are provided by two national operators (regular 
and cellular phone, and internet). The site at Rosia Montana is not very far away from the site at Baisora, so the same considerations 
apply in what concerns the major nearby universities. The site is not placed inside a reservation area, but the Apuseni National Park is 
located in the neighborhood. A medium level of chemical air pollution is generated by an open-air copper pit. This level can increase in 
the future if an envisaged gold exploitation will become operational. The level of light pollution is presently not very high. The ruins 
of the Roman stronghold Alburnus Maior are located not far away from the site, but the settlement of the observatory will not assume 
excavations in the archeological restricted area. No other cultural restrictions may apply.

\subsection{Semenic}
\label{sec4.3}

The site is located in the region of Banat, the Caras-Severin County. There is a modernized, paved road, 50 km long, connecting the site 
with the closest major city - Resita ($\approx$ 84,000 inhabitants). The closest bus station, railway station and hospital are located 
45 km away, in the city of Caransebes. The closest airport (domestic connections only) is also located in Caransebes. The closet 
international airport is located in the city of Timisoara ($\approx$ 160 km). In Figure 4 we show the road map of the area.

\begin{figure}
\resizebox{1.\hsize}{!}{\includegraphics{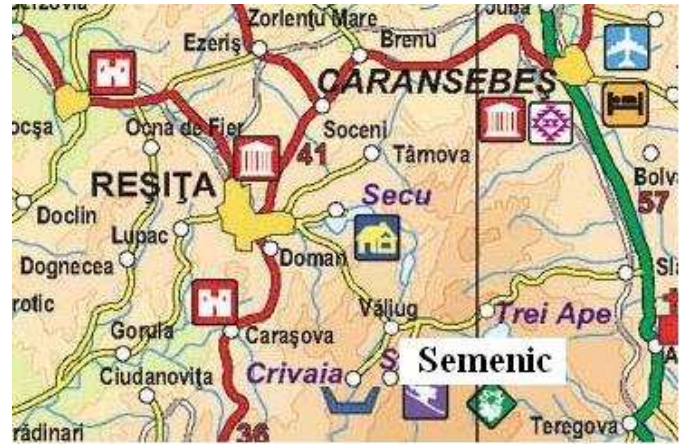} }
\caption{\label{fig4}The local road map for the site of Semenic.}
\end{figure}

Electricity and water supply are available on premises. The communications services are provided by a national operator (cellular phone 
and internet). Major technical universities in the region are located in the following cities: Resita, Timisoara and Arad ($\approx$ 200 km). 
Offices and laboratories associated to the observatory may be operated in Timisoara, the major educational center of the region. The site 
is placed inside the Semenic National Park in an area where the level of the chemical pollution is very low. The meteorological station 
whose database was used for analysis is located in the proximity of a skiing resort. Therefore, in what concerns the level of light 
pollution, the same considerations apply as in the case of Baisoara. No archeological, cultural or social restrictions may impede on the 
settlement of the observatory.

\subsection{Ceahlau}
\label{sec4.4}

The site is located in the region of Moldova, the Neamt County. The closest major city is Piatra Neamt ($\approx$ 96 km, 110,000 inhabitants). 
The closest bus station is in Durau ($\approx$ 16 km). The closest hospital and railway station are located in the city of Bicaz ($\approx$ 
66 km). The closest airport is located in the city of Bacau ($\approx$ 155 km, domestic and international connections). Rental car services 
are also available in Bacau. In Figure 5 we show the road map of the area.

\begin{figure}
\resizebox{1.\hsize}{!}{\includegraphics{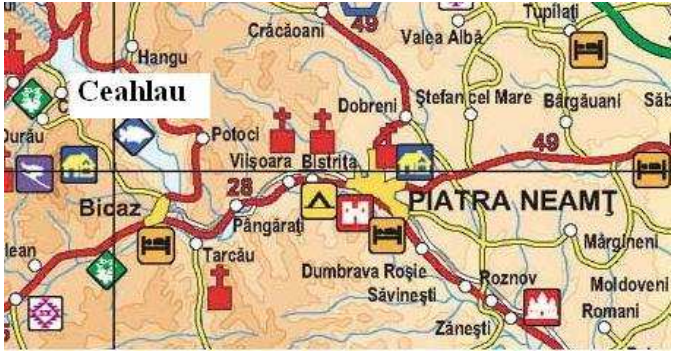} }
\caption{\label{fig5}The local road map for the site of Ceahlau.}
\end{figure} 

Electricity and water supply are available on site. The communications services are provided by a national operator (cellular phone and 
internet). Major universities with Physics and/or Mathematics departments in the region are located in Bacau and Iasi ($\approx$ 227 km). 
Offices and laboratories associated to the observatory may be operated in Iasi, the major educational center of the region. The site is 
placed inside the Ceahlau National Park. No chemical hazards can be identified in the area, but a medium level of light pollution is to 
be expected as a result of tourist activity. The darkest place in the region where to settle the telescope will be identified after in-situ 
light pollution measurements. No archeological, cultural or social restrictions may impede on the settlement of the observatory.

\section{Conclusions}
\label{sec5}

A consortium has been established to prepare the construction, in Romania, of a small, ground-based, last generation, Cherenkov telescope. 
The site testing campaign has started looking upon all places where middle altitude meteorological stations from the national network have 
been in operation over the last ten years. After local infrastructure, nearby facilities and social impact assessment criteria have been 
applied, four sites have been selected (Baisoara, Rosia Montana, Semenic, Ceahlau). All of them benefit of good local infrastructure and 
facilities. No archeological, cultural or social restrictions may impede on the settlement of the observatory at any of these sites. The 
level of chemical pollution is low. For a further refinement of the selection procedure future, in-situ light pollution measurements will 
be carried out. If none of these selected sites and their surrounding areas have a sufficiently low light background, the list of "four" 
will be upgraded with other, more isolated locations. The final choice in what concerns the site where to settle the future Romanian 
gamma-ray observatory will be made after all the tests presently underway will be completed.

\section{Acknowledgements}
\label{sec6}

The authors are grateful to Wolfgang Rhode and Michael Backes for the very enlightening discussions. This work was supported by the Romanian 
Ministry of Education, Research, Youth and Sport, the National Center for the Management of Programs (CNMP), through grant 81-010/2007/Program 4.

\end{document}